\documentclass[prd,twocolumn,floatfix]{revtex4}
\usepackage{graphicx}
\usepackage{amsmath,amssymb}
\input{epsf}
\usepackage{epsf}

\newcommand {\ga} {\ {\raise-.5ex\hbox{$\buildrel>\over\sim$}}\ }
\newcommand {\la} {\ {\raise-.5ex\hbox{$\buildrel<\over\sim$}}\ }

\begin{document}

\title{Cosmology With Non-Minimally Coupled K-Field}
\author{A.A. Sen {\footnote{anjan.ctp@jmi.ac.in}} and N.Chandrachani Devi\footnote{chandrachani@gmail.com}}
\affiliation{Center For Theoretical Physics,
Jamia Millia Islamia, New Delhi 110025, India}

\begin{abstract}
We consider a  non-minimally coupled (with gravity) scalar field with non-canonical kinetic energy. The form of the kinetic term is of Dirac-Born-Infeld (DBI) form.We study the early evolution of the universe when it is sourced only by the k-field, as well as late time evolution when both the matter and k-field are present. For the k-field, we have considered constant potential as well as potential inspired from Boundary String Field Theory (B-SFT). We show that it is possible to have an inflationary solution in early time as well as late time accelerating phases. The solutions also exhibit attractor properties in a sense that they do not depend on the initial conditions for  certain values of the parameters. 
\end{abstract}
\maketitle

\section{Introduction}
The recent rekindled interest in scalar fields, describing spin-0 particles, in the relativistic theory of gravitation stems from their potential role in solving some of the outstanding issues in cosmology. Two such issues are related to the accelerated expansion of the universe: one during its earlier evolution associated with very high energy scales and another during the much later period of evolution (more precisely at the present epoch) with much lower energy scale. In both the cases, a slowly varying energy density of a scalar field $\phi$ can mimick an effective cosmological constant. This in turn can result in the violation of the strong energy condition causing accelerated expansion.

The most widely studied scalar fields in cosmology are minimally coupled i.e the contribution from the scalar field can be separated from gravity (the Ricci scalar $R$ to be precise) and other matter fields in the action. In such realistic scalar field models describing an accelerated expansion of the universe, one requires a potential $V(\phi)$ for the scalar field which drives the acceleration. This scalar field which is responsible for the early time accelerated expansion of the universe, is called the inflaton. There are a large number of investigations regarding the role of  scalar fields in inflationary cosmology and subsequently its role in creating the initial fluctuations in the universe. We refer the reader to \cite{liddlelyth} and references therein for this purpose.

On the other hand, the discovery of the late time acceleration of the universe around the end of last century, is one of the most puzzling and daunting phenomena in recent times. First discovered by two different groups around 1999\cite{super1}, by observing  supernova type Ia explosions which are quite abundant in our universe, it has subsequently been confirmed by more recent supernova type-Ia observations by different groups\cite{super2} and indirectly confirmed by Cosmic Microwave Background Radiation (CMBR) observations\cite{cmb} and also observing the galaxy clustering by large scale redshift surveys\cite{sdss}.

To explain such phenomena, one needs to add an extra component in the energy budget of the universe, which has a large negative pressure and also dominates over all other components e.g. non-relativistic matter, radiation etc. This missing component is termed as ``dark energy'' in the literature. One optimistic choice for this dark energy can be a minimally coupled scalar field slowly rolling down its considerably flat potential \cite{scalar}. Such a scalar field is also known as ``quintessence''\cite{quint}. It is similar to the inflaton field mentioned above, but evolving in a much lower energy scale. To make the evolution of these scalar fields independent of the initial conditions, a particular class of models known as ``tracker quintessence models'' has been proposed \cite{track}, where a late time accelerating regime can be reached from a variety of initial conditions making these models more appealing. These type of scalar fields have both scaling as well as attractor property. They mimic the background matter component at early times and its late time attractor phase is accelerating which can also be stable \cite{scaling}. Tracker quintessence models with different types potential have been widely studied in recent times and their observational signatures have also been explored (see \cite{sami} for a nice review). Although minimally coupled scalar field models are interesting possibilities in modelling dark energy, one needs to do some degree of fine tuning involving the parameters of the potential.

Scalar tensor theories are generalization of the minimally coupled scalar field theories in a sense that here the scalar field is non-minimally coupled with the gravity sector of the action i.e with the Ricci scalar $R$. In these theories, the scalar field participates in the gravitational interaction, unlike its counterpart in the minimally coupled case where it behaves as a non gravitational source like any other matter field. Such a scalar partner for gravity which has a tensorial nature in general, can naturally arise in attempts at quantizing gravity or its unification with other fundamental forces in nature. Superstring theory is one such possibility where one encounters dilaton field which is non minimally coupled with the gravity sector. The reason that this non minimally coupled scalar field theory is one of the most natural alternatives to general relativity (GR) is also due to the fact that these theories respect most of the symmetries in GR like local Lorentz invariance, energy-momentum conservation etc.

Although a long range scalar field as a gravitational field was first introduced by Jordan\cite{jordan}, this kind of non minimally coupled scalar field actually attracted the attention of the researchers when Brans and Dicke invoked such a field in order to incorporate the Mach's Principle in General Theory of Relativity\cite{bd}. Brans-Dicke (BD) theory, where the scalar field is directly coupled to the Ricci scalar, has the merit of producing results which can be compared with the corresponding GR results against observations\cite{obsbd}. A further important virtue of BD theory is that it is believed to produce the GR in a particular limit(see \cite{soma} for a different result).

The possibility of having late time acceleration of the universe with a non minimally coupled scalar field ({\it a.k.a} scalar tensor theories) has also been explored in great detail\cite{stquint}. Scaling attractor solutions with non minimally coupled scalar fields have been studied with both exponential and power law potentials\cite{liddle}. Faraoni \cite{faraoni}has also studied with a non minimal coupling term $(\phi^2/2)R$ with different scalar field potentials for the late time acceleration. Bertolo {\it et al.}\cite{bertolo}, Bertolami {\it et al.}\cite{orfeu}, Ritis {\it et al.}\cite{ritis} have found tracking solutions in scalar tensor theories with different types of potential. In another work, Sen and Seshadri\cite{soma2} have obtained suitable scalar field potential to obtain power law acceleration of the universe. Saini {\it et al.}\cite{saini} and Boisseau {\it et al.}\cite{boss} have reconstructed the potential for a non minimally coupled scalar field from the luminosity-redshift relation available from the observations. Attempts have also been made to obtain an accelerating universe at present by introducing a coupling between the normal matter field and the Brans-Dicke field in a  generalised scalar tensor theory\cite{sudipta}. But the form of coupling chosen was ad-hoc and did not follow from any fundamental theory. Non-minimally coupled scalar field theories giving rise to the late time acceleration of the universe have been confronted with observational results like CMB anisotropies or growth of matter perturbations\cite{stcmb}.

In recent years, an alternative possibility of having an effective scalar field theory governed by a Lagrangian containing a non canonical kinetic term ${\cal{L}} = -V(\phi)F(X)$ where $X = (1/2)\partial_{\mu}\phi\partial^{\mu}\phi$ has been proposed. One of the most studied forms for $F(X)$ is $F(X) = \sqrt{1-2X}$. This form for the Lagrangian is called the Dirac-Born-Infeld (DBI) form. Such a model can lead to late time acceleration of the universe and popularly called ``k-essence'' in literature \cite{kess}. This field can also give rise to inflation in early universe and is called ``k-inflation''\cite{kinf}. This type of field can naturally arise in string theory\cite{asoke} and can be very interesting in cosmological context\cite{kesscos}.

In this work, we have considered non minimally coupled scalar fields having a non canonical kinetic term of DBI form. We have investigated both the early time evolution of the universe when it is dominated only by the scalar field, as well as its late time evolution when it is sourced by both matter field and scalar field. We have studied under what condition one can obtain an inflationary phase at the early times as well as an accelerating phase at late times. We have also studied the attractor properties of the solutions. We should mention that such a model was earlier considered by \cite{panda} and \cite{piao} in the context of slow-roll inflation.

\section{Non Minimally Coupled K-Field}

The action for a scalar tensor theory where a scalar field having a standard canonical kinetic term is non minimally coupled with the gravity sector is given by \cite{esposito}

\begin{equation}
S = {1\over{16\pi G}}\int d^{4}x \sqrt{-g} \left( F(\phi) R - Z(\phi) \partial_{\mu}\phi\partial^{\mu}\phi\right).
\end{equation}
Here $G$ denotes the gravitational coupling constant, $R$ is the Ricci scalar, $g$ is the determinant of the metric tensor $g_{\mu\nu}$. Here $F(\phi)$ and $Z(\phi)$ are dimensionless. $F(\phi)$ has to be positive for graviton to carry positive energy. 

We now modify the action assuming the kinetic term for the scalar field $\phi$ of DBI Form . Taking care of all  dimensional and dimensionless quantities, one can now rewrite the action as

\begin{equation}
S = {1\over{16\pi G}}\int d^{4}x \sqrt{-g} \left[ F(\phi) R - \tau_{3}V(\phi)\sqrt{1+{\eta^2} \partial_{\mu}\phi\partial^{\mu}\phi}\right].
\end{equation}
Here we have introduced two parameters $\eta$ and $\tau_{3}$ to make the action dimensionally correct. $\eta$ has the dimension of $[length]$ and $\tau_{3}$ has the dimension of $[length]^{-2}$. Varying the action (2) with respect to the metric tensor, one gets the field equation as:

\begin{eqnarray}
G_{\mu\nu}&=&-{1\over{2}}{\tau_{3} V(\phi)\over{F(\phi)}}\left[g_{\mu\nu}\sqrt{1+{\eta^{2}}\partial^{\alpha}\phi\partial_{\alpha}\phi} - {{\eta^{2}}\partial_{\mu}\phi\partial_{\nu}\phi\over{\sqrt{1+{\eta^{2}}\partial^{\alpha}\phi\partial_{\alpha}\phi}}}\right]\nonumber\\
 &+&{(F(\phi))_{,\nu ;\mu}\over{F(\phi)}} -{g_{\mu\nu}\Box F(\phi)\over{F(\phi)}}.
\end{eqnarray}
The above equations are written in so called ``Jordon Frame''. However, it is much simpler to tackle the system of equations in ``Einstein Frame'' where one can break the coupling between the scalar field $\phi$ and the Ricci scalar $R$. This can be achieved through a conformal transformation of the form $\bar{g_{\mu\nu}} = F(\phi) g_{\mu\nu}$. In this Einstein frame, the action (2) becomes
\begin{eqnarray}
S = {1\over{16\pi G}}\int d^{4}x \sqrt{-\bar{g}} [ \bar{R} - {3\over{2}}{F'(\phi)^2\over{F(\phi)}}\bar{g^{\mu\nu}}\partial_{\mu}\phi\partial_{\nu}\phi -\nonumber\\
-\tau_{3}\bar{V}(\phi)\sqrt{1+{\eta^2} F(\phi)\partial_{\mu}\phi\partial^{\mu}\phi}];
\end{eqnarray}
here $\bar{V}(\phi) = V(\phi)/F^2(\phi)$. The action is identical with that for a minimally coupled tachyon field $\phi$ with a potential $\tau_{3}\bar{V}(\phi)$  except for {\bf the extra ${3\over{2}}{F'(\phi)^2\over{F(\phi)}}\bar{g^{\mu\nu}}\partial_{\mu}\phi\partial_{\nu}\phi$ term which arises due to the non-minimal coupling}. We should point out that for a scalar field with canonical kinetic term, the action in the Einstein frame is exactly similar to the corresponding case in minimally coupled theory except the changed form of the potential which we discuss in the next paragraph. But otherwise the form of the action is exactly the same. But with a non canonical kinetic term, in the Einstein frame, one extra term gets generated. In other words, this type of non minimal coupling that we consider here, always generates a correction to ${\cal{L}}(X,\phi)$ which are of the form $A(\phi)X$, where $X = \partial^{\mu}\phi\partial_{\nu}\phi$ when transformed to the Einstein frame . This is a new feature for non minimally coupled k-essence field and can play important role in cosmological models. One point of caution is that, these models are only stable under quantum corrections as long as the non-minimal coupling term $F(\phi)$ and its variation is small.For this minimally coupled case, the DBI form of the kinetic term is protected under the five dimensional Lorentz invariance. If this extra term that appears in the minimally coupled Einstein frame, is large, then this symmetry is strongly broken and will effect significant correction in ${\cal{L}}(X,\phi)$ \cite{ref}. 

Another interesting feature is that even though one starts with a constant potential $V(\phi)=V_{0}$ in the original Jordon frame, in the Einstein frame it becomes $1/\phi^2$. This is similar to ``extended inflation'' scenario\cite{la}, where the constant potential for the inflaton field in its false vacuum state, gets modified to exponential potential due to the conformal transformation. This results much slower power law inflation for the universe solving some of the problems present in the original Guth's model of inflation\cite{guth}. Although this model has its own  problem, non minimal coupling and subsequent conformal transformation play a crucial role for the initial success of this model.

Next we assume a homogeneous and isotropic universe described by a Friedmann-Robertson-Walker (FRW) spacetime. It is given by the spacetime metric
\begin{equation}
ds^{2} = - dt^2 + a^{2}(t) \left[dr^2 +r^2 d\theta^2 + r^2 Sin^2 \theta d\chi^2\right]
\end{equation}
where $a(t)$ is scale factor of the universe. One also has to assume the form for $F(\phi)$ for further calculations. One of the most studied  forms in the literature is $F(\phi) = \phi$ which is also called the Brans-Dicke (BD) form\cite{bd}. The original motivation for the BD theory was to implement the Mach principle in relativistic theory of gravity. This was done by promoting the Newton's constant to the role of a dynamical field determined by the environment. Later on, it was shown that the low energy limit of the bosonic string theory is indeed similar to the BD theory for some particular parameter value\cite{callan,fradkin}. Moreover, the higher dimensional Kaluza-Klein gravity can give rise to BD theory after dimensional reduction. In our subsequent calculations we shall assume $F(\phi) = \phi$.
\begin{figure}[t]
\centerline{\epsfxsize=3.2truein\epsfbox{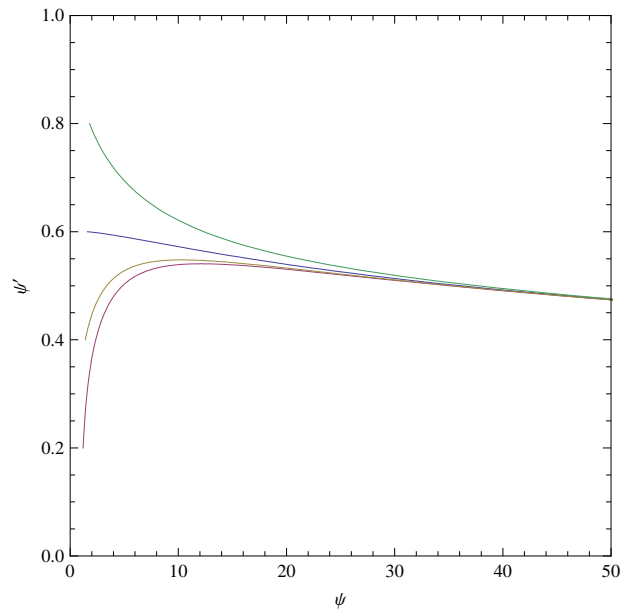}}
\caption{Phase portrait in the $\psi^{'}-\psi$ phase plane for constant potential. The different initial conditions for $(\psi^{'}_{i},\psi_{i})$ are (1.8,0.8), (1.6,0.6), (1.4,0.4) and (1.2,0.2) respectively from top to bottom. $\alpha$ = 0.2 for this plot.}
\end{figure}

\begin{figure}[t]
\centerline{\epsfxsize=3.2truein\epsfbox{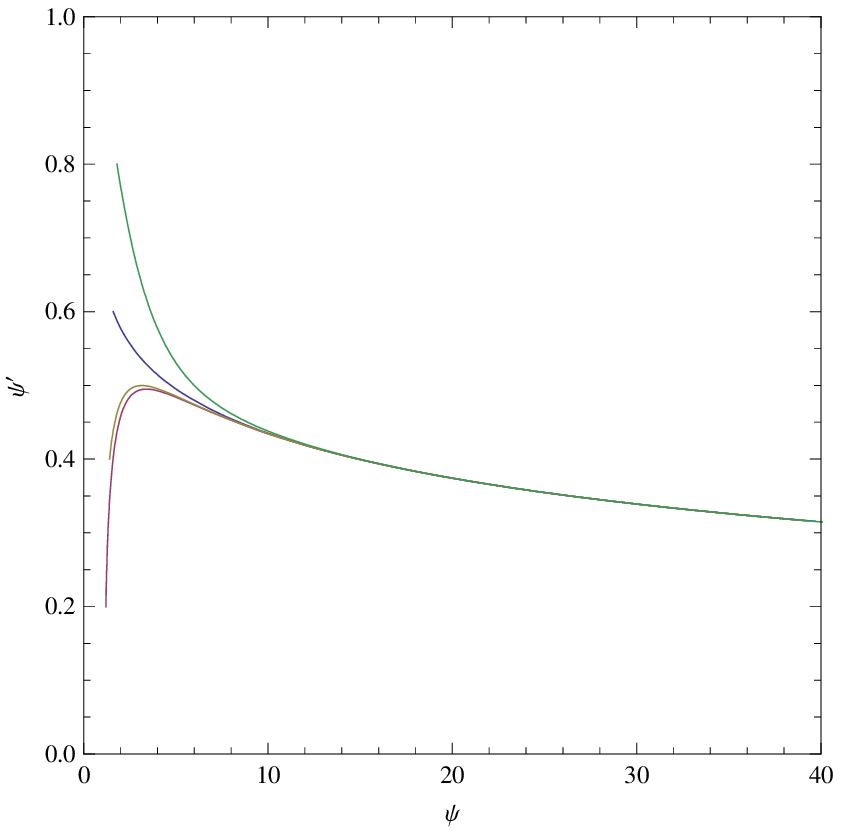}}
\caption{Phase portrait in the $\psi^{'}-\psi$ phase plane for constant potential. The different initial conditions for $(\psi^{'}_{i},\psi_{i})$ are (1.8,0.8), (1.6,0.6), (1.4,0.4) and (1.2,0.2) respectively from top to bottom. $\alpha$ = 0.8 for this plot.}
\end{figure}

With these choices, Einstein equations derived from the action (4) becomes,
\begin{eqnarray}
3\bar{H}^{2} = {\tau_{3} V(\phi)\over{2\phi^2 \sqrt{1-\eta^2\phi \phi^{'2}}}} + {3\over{4}}{\phi^{'2}\over{\phi^2}}\\
2\bar{H^{'}} + 3\bar{H}^2 = {\tau_{3} V(\phi)\over{2\phi^2}}\sqrt{1-\eta^2\phi \phi^{'2}} - {3\over{4}}{\phi^{'2}\over{\phi^2}},
\end{eqnarray}
where prime denotes differentiation w.r.t $\bar{t}$, the time in Einstein frame and $\bar{H} = {d\over{d\bar{t}}}\log\bar{a}$. We have also assumed $16\pi G = 1$. 

One can further redefine the scalar field to have a more simpler equation. This is done by defining $\eta^{2} \phi \phi^{'2} = \psi^{'2}$. Using this, the field equations (10) and (11) now become
\begin{eqnarray}
3\bar{H}^{2} = {V(\psi)\alpha\over{\psi^{4/3} \sqrt{1- \psi^{'2}}}} + {1\over{3}}{\psi^{'2}\over{\psi^2}}\\
2\bar{H^{'}} + 3\bar{H}^2 = {V(\psi)\alpha\over{\psi^{4/3}}}\sqrt{1-\psi^{'2}} - {1\over{3}}{\psi^{'2}\over{\psi^2}},
\end{eqnarray}
where $\alpha = [({2\over{3}}\eta)^{4/3} \tau_{3}]/2$.  One can also combine the above two equations to get one equation for the evolution of the scalar field $\psi$:

\begin{eqnarray}
\left[{\alpha V_{1}(\psi)\over{(1-\psi^{'2})^{3/2}}}+{2\over{3}}{1\over{\psi^2}}\right]\psi^{''} &+&\psi^{'}H\left[{2\over{\psi^{2}}}+{3\alpha V_{1}(\psi)\over{\sqrt{1-\psi^{'2}}}}\right]\nonumber\\
-{2\over{3}}{\psi^{'2}\over{\psi^{3}}}+{\alpha V_{1}^{'}(\psi)\over{\sqrt{1-\psi^{'2}}}} &=& 0,
\end{eqnarray}
where $V_{1}(\psi) ={ V(\psi)\over{\psi^{4/3}}}$ and $V_{1}^{'}(\psi) = {dV_{1}(\psi)\over{d\psi}}$.
 Here also prime denotes differentiation w.r.t the time in Einstein frame. We now consider different forms for the potential to study the evolution of the universe. 

\subsection{$V(\phi)$ = constant= $V_{0}$}

In this case, as
we  are not considering any other field (either radiation or matter),and  the universe only contains the scalar
 field $\psi$, we are primarily interested in the early universe scenario, more specifically we look for possible
  inflationary scenario in this model.
\begin{figure}[t]
\centerline{\epsfxsize=3.2truein\epsfbox{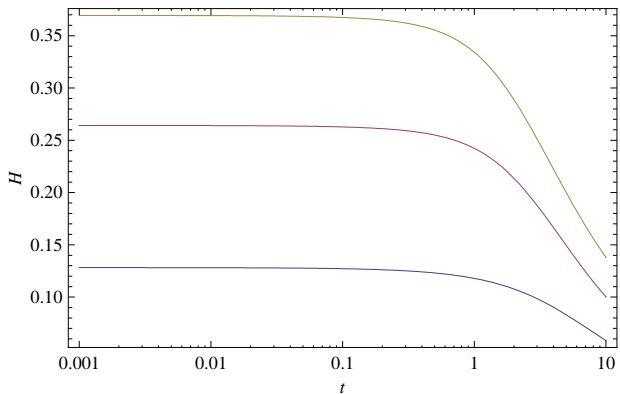}}
\caption{Variation of the Hubble parameter $H(t)$ with time for $V(\phi)$=constant. We have choosen $\alpha = 1,0.5,0.1$ respectively from top to bottom.}
\end{figure}

The potential becomes $V_{1}(\psi) = V_{0}/\psi^{4/3}$ in eqn (14). We assume the constant $V_{0} = 1$ in our subsequent calculations without any loss of generality. With this choice for the potential, we now have the system of equations (12), (13) and (14). We first investigate whether the solution for this system of equations is sensitive to the initial condition. More specifically, we check whether the evolution of the scalar field in the $\psi^{'}-\psi$ phase-plane has any attractor behaviour. Here $\psi^{'} = {d\psi\over{d{\bar{t}}}}$. Similar kind of study has been done earlier for minimally coupled fields with DBI type kinetic term\cite{samicop}. One can see from equation (12) that we only need the initial conditions on $\psi$ and $\psi^{'}$. The initial condition on $H$ is fixed through equation (12). In figure 1 and figure 2, we have shown the evolution in the $\psi^{'}-\psi$ phase plane with two different values for the parameter $\alpha$. In both cases, it shows that the evolution of the scalar field does not depend on the initial conditions. It always has a late time attractor. This feature is crucial in order to have a reliable inflationary model whose predictions should not depend on the initial conditions.
\begin{figure}[t]
\centerline{\epsfxsize=3.2truein\epsfbox{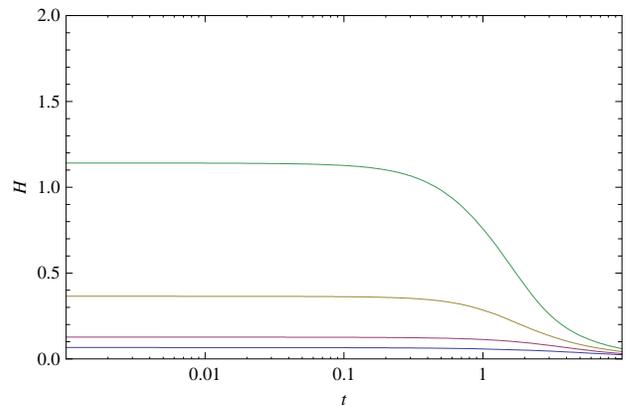}}
\caption{Variation of the Hubble parameter $H(t)$ with time for $V(\phi)=exp(-\phi^{2}/4)$. We have choosen $\eta = 1$ and $\tau_{3} = 1,10,100$ from bottom to top respectively.}
\end{figure}
Next we show the behaviour of the Hubble parameter $H(t)$ for different values of the parameter $\alpha$ in figure 3. One can see for all the cases, it starts with an inflationary era (showing $H(t)$ = constant) and then exits from this inflationary phase. The time of exit depends on the parameter $\alpha$ in a sense that lower values of $\alpha$ causes the exit at later time. So one can tune the parameter $\alpha$ to get sufficient number of e-folding to solve the horizon or flatness problem.

\subsection{$V(\phi) = exp(-\phi^2/4)$}

Next we consider the potential $V(\phi) = exp(-\phi^2/4)$ in the action (2). This type of Gaussian potential can arise in boundary string field theory (B-SFT)\cite{bsft} and has been studied for models of inflation driven by minimally coupled tachyon field. In this case the potential $V(\psi)$ in Einstein frame becomes $V(\psi) = exp(-A \psi^{4/3}/4)$, where $A= \tau_{3}/\alpha$. We have chosen $\eta =1$ throughout for this case. Similar to the constant potential case, described above, one can study the variation of the Hubble parameter with time. This has been shown in figure 4. Here also, the universe undergoes an exponential expansion in the early time, and then it exits from this inflationary era. The duration for which this rapid expansion lasts, depends on the parameter $\tau_{3}$ with lower values of $\tau_{3}$ results larger duration for inflation.

We have also studied the phase portrait in the $\psi^{'}-\psi$ plane to see the attractor property. This has been shown in figure 5. Here we have assumed $\tau_{3} = 10$. As in the constant potential case, here also we have the attractor property so that the late time solution does not depend on the initial conditions. But this attractor property is  lost as one decreases the value of the $\tau_{3}$ parameter. This has been shown in figure 6 where we have assumed $\tau_{3} = 1$. Hence for this potential $\tau_{3}$ plays a crucial role for the attractor solution. As we mention before that we have assumed $\eta=1$ for this case. But with higher values of $\eta$ one can get attractor behaviour with smaller $\tau_{3}$.

\begin{figure}[t]
\centerline{\epsfxsize=3.2truein\epsfbox{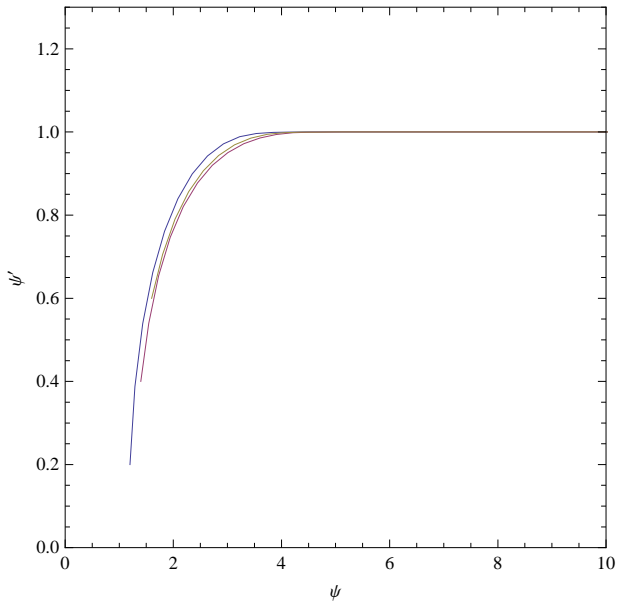}}
\caption{Phase portrait in the $\psi^{'}-\psi$ phase plane for $V(\phi) = exp(-\phi^2/4)$. The different initial conditions for $(\psi^{'}_{i},\psi_{i})$ are (1.6,0.6), (1.4,0.4)and  (1.2,0.2) respectively from top to bottom. $\eta$ = 1 and $\tau_{3} = 10$ for this plot.}
\end{figure}
\begin{figure}[t]
\centerline{\epsfxsize=3.2truein\epsfbox{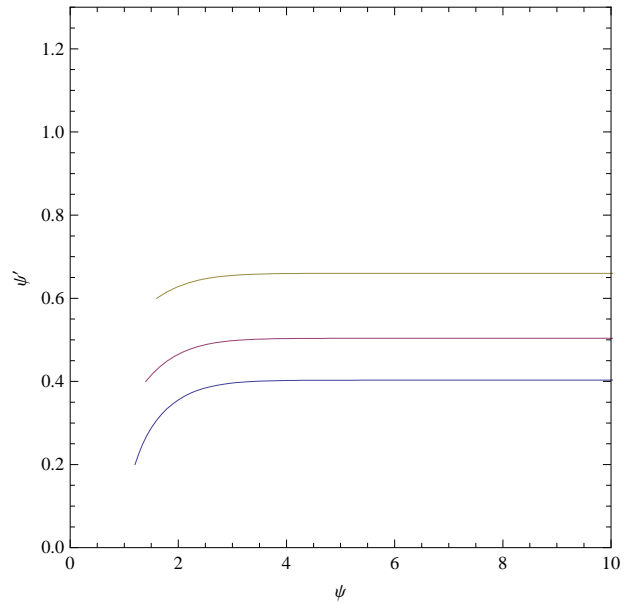}}
\caption{Same as figure 5 except $\tau_{3} = 1$ in this plot.}
\end{figure}

One can compare this non-minimally coupled case with the corresponding minimally coupled case with the same exponential potential. This has been studied by Sen \cite{asoke} and Frolov {\it et al.} \cite{frolov}. In the minimally coupled case, $\psi^{'}$ approaches 1 asymptotically and the k-field behaves as a pressure-less dust. We study the corresponding situation in the non-minimally coupled case. The behaviour of $\psi^{'}$ is shown in figure 7. Here also we have assumed $\eta=1$. But higher values of $\tau_{3}$ which is also necessary for attractor behaviour, the figure shows that here also $\psi^{'}$ approaches 1 very quickly. But there is an extra term in r.h.s of equation (12) which arises due to non-minimal coupling. But still the asymptotically the pressure vanishes, similar to the minimally coupled case. This has been shown in figure 8.

\section{Non minimal K-essence}
In this section we consider the late time solution for the universe where it is dominated by a non-relativistic matter together with a non canonical scalar field which is non minimally coupled with the gravity sector. We call it ``non minimal k-essence''.

There are different approaches for adding matter in scalar tensor theories. This is because of the ambiguity of representing the physical space-time structure: whether it belongs to the Jordan frame or to Einstein Frame. And this ambiguity has an elaborate representation in the literature dealing with scalar tensor theories \cite{barrow,gr,spin2,quant,third,brans}. Choosing a physical frame is utmost important for theoretical investigations. Also the physical metric should be singled out for the pure gravity only vacuum case and subsequently the matter field should be added. In a detail study, Magnano and Sokolowski\cite{magnano} have argued in favour of Einstein frame as a physical frame in scalar tensor theories.  They showed that if a Lagrangian in scalar tensor theory admits a flat space as a stable solution, then it is always equivalent to a scalar field with potential minimally coupled to conformally transformed metric. Then any matter should be minimally coupled with the transformed metric and not with the scalar field. But this does not apply to theories where the coupling between the matter field and scalar field already exists in more fundamental theory. As an example, in string theory, the four dimensional effective action contains a number bosonic fields which couple both with the metric and the dilaton.
\begin{figure}[t]
\centerline{\epsfxsize=3.2truein\epsfbox{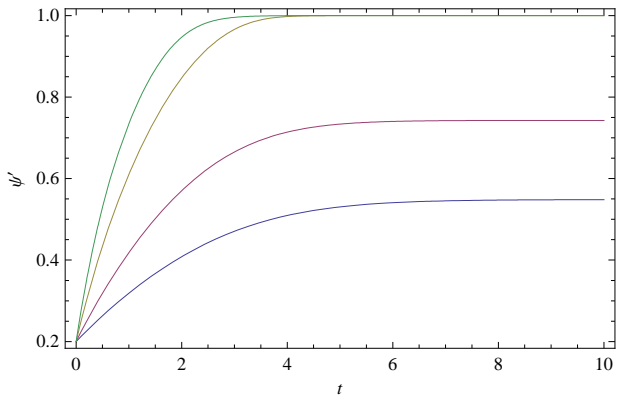}}
\caption{Behaviour of $\psi^{'}$ with time. $\eta=1$. $\tau_{3} = 1,2,5,10$ from bottom to top.}
\end{figure}
\begin{figure}[t]
\centerline{\epsfxsize=3.2truein\epsfbox{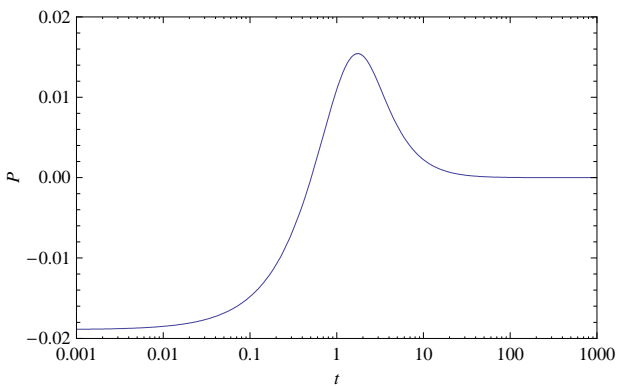}}
\caption{Behaviour of pressure $P$ for the potential $V(\phi)=exp(-\phi^{2}/4)$ with time. $\eta=0.5$ and $\tau_{3} = 10$ .}
\end{figure}
In our subsequent calculations, we consider both the Einstein frame and Jordan frame for pure gravity case and add  matter fields in these frames and compare our results. For this purpose, we consider only $V(\phi)= V_{0}$ case. But one can consider any potential in principle.

\subsection{Adding Matter in Jordan Frame}

In this case we add matter in the Jordan Frame, i.e we add the matter in action (2):
\begin{eqnarray}
S &=& {1\over{16\pi G}}\int d^{4}x \sqrt{-g} \left[ F(\phi) R - \tau_{3}V(\phi)\sqrt{1+{\eta^2} \partial_{\mu}\phi\partial^{\mu}\phi}\right]\nonumber\\
 &+& S_{matter}.
\end{eqnarray}

\begin{figure}[t]
\centerline{\epsfxsize=3.2truein\epsfbox{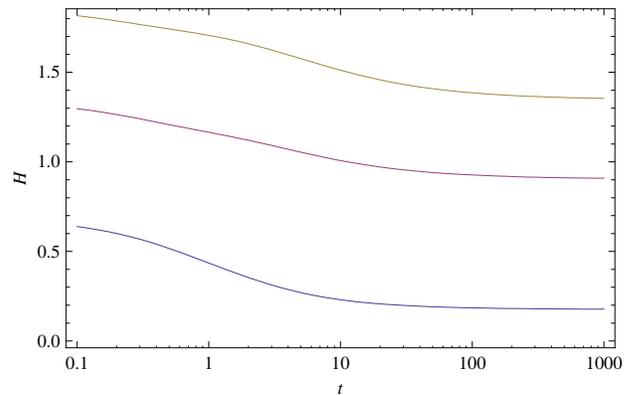}}
\caption{Variation of the Hubble parameter $H(t)$ with time for $V(\phi)$=constant plus matter. We have chosen $\alpha = 1,5,10$ respectively from bottom to top.}
\end{figure} 
We have considered pressure-less dust for our purpose. We have also transformed our equations in Einstein Frame and they take the form (assuming $F(\phi)=\phi$):
\begin{eqnarray}
3\bar{H}^{2} = {8\pi G {\bar{\rho}}}+{V(\psi)\alpha\over{\psi^{4/3} \sqrt{1- \psi^{'2}}}} + {1\over{3}}{\psi^{'2}\over{\psi^2}}\\
2\bar{H^{'}} + 3\bar{H}^2 = {V(\psi)\alpha\over{\psi^{4/3}}}\sqrt{1-\psi^{'2}} - {1\over{3}}{\psi^{'2}\over{\psi^2}}
\end{eqnarray}
 \begin{figure}[t]
\centerline{\epsfxsize=3.2truein\epsfbox{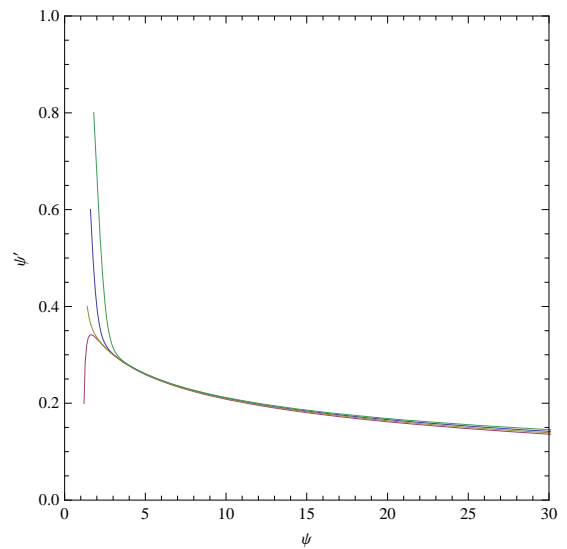}}
\caption{Phase portrait in the $\psi^{'}-\psi$ phase plane for $V(\phi) = constant$ and with matter included in Jordan frame. The different initial conditions for $(\psi^{'}_{i},\psi_{i})$ are (1.8,0.8),(1.6,0.6), (1.4,0.4)and  (1.2,0.2) respectively from top to bottom. $\alpha$ = 5.}
\end{figure} 
where $\bar{\rho}$ is the energy density in the Einstein frame. The field $\psi$ and the constant $\alpha$ are defined section II. The equation of motion for the scalar field and the energy conservation equation for matter now become:
\begin{eqnarray}
\left[{\alpha V_{1}(\psi)\over{(1-\psi^{'2})^{3/2}}}+{2\over{3}}{1\over{\psi^2}}\right]\psi^{''} &+&\psi^{'}H\left[{2\over{\psi^{2}}}+{3\alpha V_{1}(\psi)\over{\sqrt{1-\psi^{'2}}}}\right]\nonumber\\
-{2\over{3}}{\psi^{'2}\over{\psi^{3}}}+{\alpha V_{1}^{'}(\psi)\over{\sqrt{1-\psi^{'2}}}} &=& {1\over{3}}{{\bar{\rho}}\over{\psi}},
\end{eqnarray}

\begin{equation}
{\bar{\rho}}^{'} + 3{\bar{H}}{\bar{\rho}} = - {1\over{3}}{{\bar{\rho}}\psi^{'}\over{\psi}}
\end{equation}
 \begin{figure}[t]
\centerline{\epsfxsize=3.2truein\epsfbox{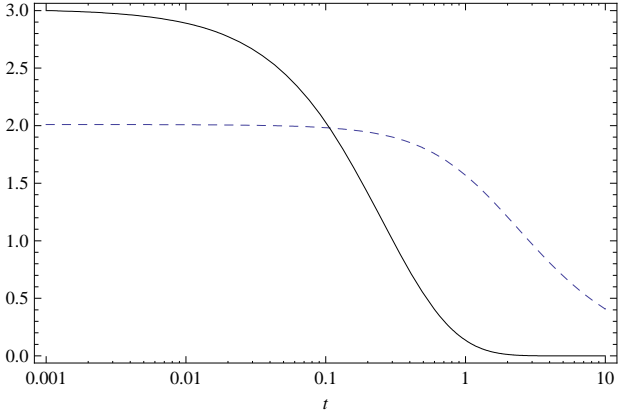}}
\caption{Behaviour of the energy densities (solid for $\rho_{matter}$ and dashed for $\rho_{\psi}$) for $V(\phi) = constant$ case. $\alpha$ = 5.}
\end{figure} 
One can notice that in this frame there is a coupling between the matter ${\bar{\rho}}$ and the scalar field $\psi$. This is usual for the case where one incorporates matter in the Jordan frame.  In figure 9 we show the behaviour of the Hubble parameter $H(t)$ as function of time for different values of the $\alpha$ parameter. It shows that a late time acceleration phase ($H(t) = constant$) exists in this model. The epoch when this accelerating phase starts, depends on the $\alpha$ parameter.
\begin{figure}[t]
\centerline{\epsfxsize=3.7truein\epsfbox{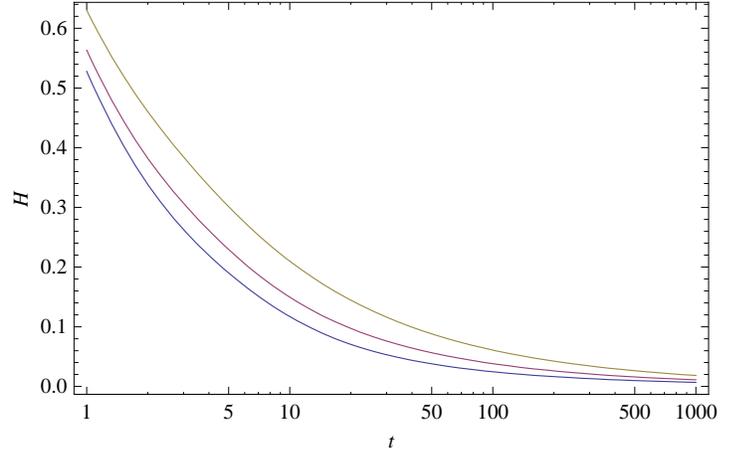}}
\caption{Variation of the Hubble parameter $H(t)$ with time for $V(\phi)$=constant adding matter in Einstein frame. We have chosen $\alpha = 0.5,1,2$ respectively from bottom to top.}
\end{figure} 
\begin{figure}[t]
\centerline{\epsfxsize=3.2truein\epsfbox{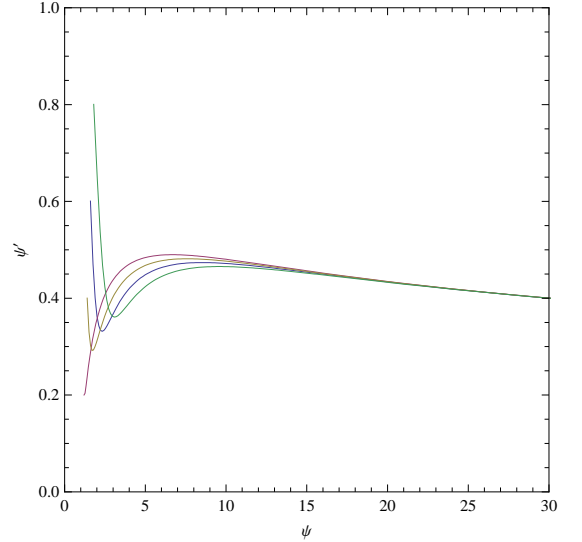}}
\caption{Phase portrait in the $\psi^{'}-\psi$ phase plane for $V(\phi) = constant$ and with matter included in Einstein frame. The different initial conditions for $(\psi^{'}_{i},\psi_{i})$ are (1.8,0.8),(1.6,0.6), (1.4,0.4)and  (1.2,0.2) respectively from top to bottom. $\alpha$ = 5.}
\end{figure} 
We also study the phase diagram in $\psi^{'}-\psi$ plane, to check whether our solution depends on the initial condition. We show the result in figure 10. It is shown for different initial conditions with $\alpha = 5$. We have showed the behaviour of the energy densities for the matter and scalar field in figure 11 for $\alpha = 5$. It shows that although matter dominates the energy budget at early time, but eventually the scalar field takes over and starts dominating. 

\subsection{Adding Matter in Einstein Frame}
\vspace{2mm}
Next we add matter in the Einstein frame i.e we consider the physical metric for pure gravity for vacuum in Einstein frame and then add matter in it. The action now becomes:
\begin{eqnarray}
S = {1\over{16\pi G}}\int d^{4}x \sqrt{-\bar{g}} [ \bar{R} - {3\over{2}}{F'(\phi)^2\over{F(\phi)}}\bar{g^{\mu\nu}}\partial_{\mu}\phi\partial_{\nu}\phi -\nonumber\\
-\tau_{3}\bar{V}(\phi)\sqrt{1+{\eta^2} F(\phi)\partial_{\mu}\phi\partial^{\mu}\phi}] + S_{matter}
\end{eqnarray}

 We write the field equations in this case and they take the form (assuming $F(\phi) = \phi$):
\begin{eqnarray}
3\bar{H}^{2} = {8\pi G \rho}+{V(\psi)\alpha\over{\psi^{4/3} \sqrt{1- \psi^{'2}}}} + {1\over{3}}{\psi^{'2}\over{\psi^2}}\\
2\bar{H^{'}} + 3\bar{H}^2 = {V(\psi)\alpha\over{\psi^{4/3}}}\sqrt{1-\psi^{'2}} - {1\over{3}}{\psi^{'2}\over{\psi^2}}.
\end{eqnarray}
 
The field $\psi$ and the constant $\alpha$ are defined in section II. The conservation equation for matter field is given by
\begin{equation}
{{\rho}}^{'} + 3{\bar{H}}{{\rho}} = 0
\end{equation}
\begin{figure}[t]
\centerline{\epsfxsize=3.2truein\epsfbox{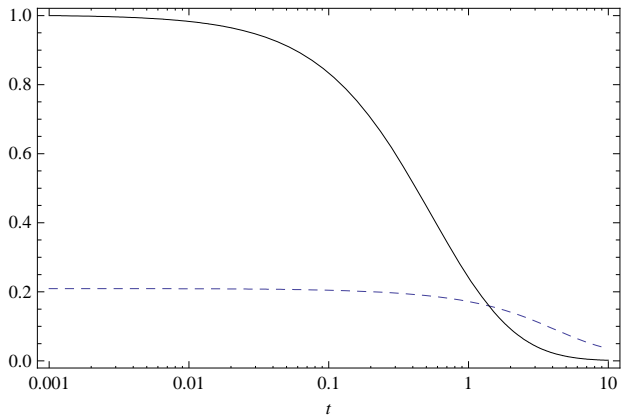}}
\caption{Behaviour of the energy densities (solid for $\rho_{matter}$ and dashed for $\rho_{\psi}$) for $V(\phi) = constant$ case with matter added in Einstein frame. $\alpha$ = 0.5.}
\end{figure}
We have shown the behaviour of the Hubble parameter $H(t)$ as a function of time in figure 12. Here also one can get a late time accelerating period (constant $H(t)$). But one needs to have low $\alpha$ to get the acceleration, in comparison to the previous case where we add matter in Jordan frame. We also study the phase portrait in $\psi^{'}-\psi$ plane in figure 13. It also shows the attractor behaviour. In this case the range for $\alpha$ for which one gets the attractor property is wider than the previous one. The behaviours of the two energy densities are also shown in figure 14.  In this case also, although the matter dominates in the early time, the scalar field starts dominating in the late time resulting acceleration.
\section{Conclusion}
In this work, we have considered a k-field having DBI-type kinetic term, which is  non minimally coupled with gravity. The coupling term is of BD type i.e $F(\phi) = \phi$ in the action. We have investigated in detail the cosmological solutions in this model. Inflationary solutions in D-Brane models with the scalar field having a DBI-type kinetic term has been previously investigated by Underwood \cite{underwood}. 

We have considered both the early era for universe evolution, when it is sourced only by the scalar field as well as its late time evolution when both scalar and the matter fields are present. Our main aim is to check, whether it is possible to have a epoch of accelerated expansion either in early time or in late time. Our results show that with proper choice of parameters, it is possible to have both inflation and late time acceleration in this model. More over we have shown that these solutions show attractor property in a sense that they do not depend on the initial conditions. For the inflation case, we have considered both the constant potential and potential inspired from the BSFT theories, in our original action in Jordan frame. But in the Einstein frame, where we study the solutions, the potentials get modified. Moreover the r.h.s of the field equations (eqns 10 and 11), contains an extra term which is due to the non minimal coupling. This term  is completely new and does not occur in the corresponding minimally coupled case. It can play an important role specially for models where one can have a fast roll inflationary phase. 

As our results show, this model can have interesting cosmological implications. We have shown that the model allows an inflationary phase. One has to now apply it to a more realistic inflationary scenario (so called slow-roll model), and study the implications, e.g. whether it allows sufficient e-folding, what is the power spectrum for this k-field and whether that matches with the observations. Scalar field with DBI type kinetic term minimally coupled with gravity has already been considered for possible inflaton field\cite{kinf1}. But such models with specific potentials arising from realistic string theory models has been problematic in inflationary model building\cite{kinfprob}. But introducing non-minimal coupling with gravity may alleviate such problems.

Also although we have shown that attractor solutions exist in this model for both early time and late time evolution, a more detail study is required to study the fixed points, and their stability.

We have shown that the model also allows a late time accelerating phase, when the universe is sourced by matter together with the k-field. One has to now make a detailed study regarding the observational aspects of this model.

To summarise, non-minimally coupled k-field is an attractive model for cosmological purpose for both inflation and late time accelerating universe. But further study is needed to confirm its viability.

{\it note added:} Just before submitting the paper in arXiv, another paper with similar model has appeared \cite{recent}. The authors have investigated similar type of non-minimally coupled k-field models but in a different approach.

\vspace{5mm}  
\section{Acknowledgement}
The authors acknowledge the financial support provided by the University Grants Commission, Govt. Of India, through the major research project grant (Grant No:  33-28/2007(SR)).


\begin{thebibliography}{99}
\bibitem{liddlelyth}A.R. Liddle and D.H. Lyth {\it Cosmological Inflation and Large Scale Structure}, Cambridge University Press (2000).
\bibitem{super1}S. Perlmutter et al. Ap.J., {\bf 565} (1999);
A.G. Riess et al., Astron.J., {\bf 116}, 1009 (1999).
\bibitem{super2}R.A. Knop et al., Ap.J., {\bf 598}, 102 (2003);
A.G. Riess et al., Ap.J., {\bf 607}, 665 (2004);
T.M. Davis et al., Ap.j., {\bf 666}, 716 (2007).
\bibitem{cmb}E. Komatsu et al., arXive:0803.0570 [astro-ph];
J. Dunkley et al., arXive:0803.0586 [astro-ph].
\bibitem{sdss}D.J. Eisenstein et al. Ap.J, {\bf 633}, 560 (2005).
\bibitem{scalar}B. Ratra and P.J.E. Peebles, Phys.Rev.D, {\bf 37}, 3406 (1988);
C. Wetterich, Nucl.Phys.B, {\bf 302}, 668 (1988).
\bibitem{quint}E.J. Copeland, A. Liddle and D. Wands, Phys.Rev.D, {\bf 57}, 4686 (1998);
P.G. Ferreira and M. Joyce, Phys.Rev.D, {\bf 58}, 023503 (1998).
\bibitem{track}R.R. Caldwell, R. Dave and P.J. Steinhardt, Phys.Rev.Lett, {\bf 80}, 1582 (1988);
I. Zlatev L.M. Wang and P.J. Steinhardt, Phys.Rev.Lett., {\bf 82}, 896 (1999);
P.J. Steinhardt, L.M. Wang and I Zlatev, Phys.Rev.D., {\bf 59}, 123504 (1999).
\bibitem{scaling}A.R. Liddle and R.J. Scherrer, Phys.Rev. D {\bf 59}, 023509 (1999);
T. Barreiro, B.de Carlos and E.J. Copeland, Phys.Rev.D, {\bf 58}, 083513 (1998);G.Huey et al. Phys.Lett.B, {\bf 476}, 379 (2000);
A.de la Macorra and G. Piccinelli, Phys.Rev.D, {\bf 61}, 123503 (2000);
E.J. Copeland A. Liddle and D. Wands, Phys.Rev.D, {\bf 57}, 4686 (1998);
P.G. Ferreira and M. Joyce, Phys.Rev.D., {\bf 58}, 023503 (1998);
S.Carroll, Phys.Rev.Lett., {\bf 81}, 3067 (1998);
L.Wang et al., Astrophys.J., {\bf 530}, 17 (2000);
P.Brax and J. Martin, Phys.Lett, {\bf B 468}, 40 (1999);
V.Sahni and A. Starobinsky, Int.J.Mod.Phys.D, {\bf 9}, 373 (2000);
P.Brax and J.Martin, Phys.Rev.D., {\bf 61}, 103502 (2000).
\bibitem{sami}T.Padmanabhan Phys.Rept., {\bf 380}, 235 (2003);
E.J.Copeland, M.Sami and S.Tsujikawa, Int. J. Mod. Phys. D., {\bf 15}, 1753 (2006).
\bibitem{jordan}P. Jordan, Z.Phys,{\bf 157} 112 (1959).
\bibitem{bd}C. Brans and R.H. Dicke, Phys.Rev.D, {\bf 124}, 925 (1961).
\bibitem{obsbd}T. Damour and G. Esposito-Farese, Class.Quant.Grav., {\bf 9}, 2093 (1992);T. Damour and G.Esposito-Farese, Phys.Rev.D, {\bf 53}, 5541 (1996);
 T. Damour and G.Esposito-Farese, Phys.Rev.D, {\bf 54}, 1474 (1996);
 C.M. Will, Phys.Rev.D, {\bf 50}, 6058 (1994);
T. Damour and G.Esposito-Farese, Phys.Rev.D, {\bf 58}, 042001 (1998);
T.Damour and K. Nordtvedt Phys.Rev.Lett, {\bf 70}, 2217 (1993).
\bibitem{soma}N.Banerjee and S. Sen, {\bf 56}, 1334 (1997).
\bibitem{ref} We thank the anonymous referee for pointing out this. 
\bibitem{guth}A. Guth, Phys. Rev. Lett., {\bf 49}, 1110 (1982).
\bibitem{callan}C.G. Callan, D. Friedan, E.J. Martinez and M.J.Perry, Nucl.Phys., {\bf B 262}. 593 (1985).
\bibitem{fradkin}E.S. Fradkin and A.A. Tseytlin, Nucl.Phys., {\bf B 261}, 1 (1985).
\bibitem{la}D. La and P.J. Steinhardt, Phys.Rev.Lett., {\bf 62}, 376 (1989).
\bibitem{stquint}N. Banerjee and D. Pavon, Class.Quant.Grav., {\bf 18}, 593 (2001);
A.A. Sen and S. Sen, Mod.Phys.Lett.A, {\bf 16}, 1303 (2001);
S. Sen and A.A. Sen, Phys. Rev.D {\bf 63}, 124006 (2001);
J.P. Uzan, Phys.Rev.D, {\bf 59}, 123510, (1999);
T. Chiba, Phys.Rev.D {\bf 60}, 083508 (1999);
F. Perrotta, C. Baccigalupi and S. Matarrese, astro-ph/9906066;
\bibitem{liddle}A.R. Liddle and R.J. Scherrer, Phys.Rev.D {\bf 59}, 023509 (1998).
 \bibitem{faraoni}V. Faraoni, gr-qc/0002091.
\bibitem{bertolo}N. Bertolo and M. Piertroni, hep-th/9908521.
\bibitem{orfeu}O. Bertolami and P.J. Martins, Phys.Rev.D, {\bf 61}, 064007 (2000).
\bibitem{ritis}R. de Ritis, A.A. Marino, C. Rubano and P. Scudellaro, Phys.Rev.D., {\bf 62}, 043506 (2000).
\bibitem{soma2}S. Sen and T.R. Seshadi, gr-qc/0007079.
\bibitem{saini}T.D. Saini, S. Raychaudhury, V. Sahni and A.A. Starobinsky, Phys.Rev.Lett., {\bf 85}, 1162 (2000).
\bibitem{boss}B. Boisseau, G. Esposito-Farese, D.Ploarski and A.A. Starobinski, Phys.Rev.Lett., {\bf 85}, 2236 (2000).
\bibitem{sudipta}S. Das and N. Banerjee, Gen.Rel.Grav., {\bf 38}, 785 (2006).
\bibitem{stcmb}E. Gaztanaga and J.A. Lobo, astro-ph/0003129;
F. Perrotta, C. Baccigalupi and S. Matarrese, astro-ph/9906066;
X. Chen and M. Kamionkowski, Phys.Rev.D, {\bf 60}, 104036 (1999);
D.J.Holden and D. Wands, gr-qc/9908026;
T. Damour and B. Pichon, Phys.Rev.D {\bf 59}, 123502 (1999).
\bibitem{kess}C. Armendariz-Picon, V. Mukhanov and P.J. Steinhardt, Phys.Rev.D, {\bf 63}, 103510 (2001);
C. Armendariz-Picon, V. Mukhanov and P.J. Steinhardt, Phys.Rev.Lett, {\bf 85}, 4438 (2000);
T. Chiba, Phys.Rev.D, {\bf 66}, 063514 (2002).
\bibitem{kinf} Armendariz-Picon, T. Damour and V. Mukhanov, Phys.Lett.B, {\bf 458}, 209 (1999).
\bibitem{asoke}A. Sen, JHEP, {\bf 0204}, 048 (2002);
A. Sen, JHEP, {\bf 0207}, 065 (2002).
\bibitem{samicop}E.J.Copeland, M.R. Garousi, M.Sami and S. Tsujikawa, Phys.Rev.D., {\bf 71}, 043003 (2005).
\bibitem{frolov}A. Frolov, L. Kofman and A. Starobinsky, hep-th/0204187.
\bibitem{kesscos}J.S. Bagla, H.K. Jassal and T Padmanabhan, Phys.Rev.D, {\bf 67}, 063504 (2003);
J.M. Aguirregabiria and R. Lazkoz, Phys.Rev.d, {\bf 69}, 123502 (2004);
A.Das, S. Gupta, T.D. Saini and S. Kar, Phys.Rev.D, {\bf 72}, 043528 (2005).
\bibitem{panda}P. Chingangbam, S. Panda and A. Deshamukhya, JHEP, {\bf 0502}, 052 (2005).
\bibitem{piao}Y-S. Piao, Q-G. Huang, X. Zhang and Y-Z, Zhang, Phys. Lett., {\bf B 570}, 1 (2003).
\bibitem{esposito}G.Esposito-Farese and D.Polarski, Phys.Rev.D, {\bf 63}, 063504 (2001).
\bibitem{bsft}D.Kutasov, M. Marino and G.W. Moore, JHEP, {\bf 0010}, 045 (2000).
\bibitem{barrow}K. Maeda, Phys.Rev.D, {\bf 36}, 858 (1988);
J.D. Barrow and K. Maeda, Nucl. Phys., {\bf B 341}, 294 (1990);
T. Damour, G. Gibbons and C. Gundlach, Phys.Rev.Lett., {\bf 64}, 123 (1990);
J.D. Barrow, Phys.Rev.D., {\bf 47}, 1475 (1993);
T. Damour and C. Gundlach, Phys.Rev.D, {\bf 43}, 3873 (1991).
\bibitem{gr}G.W. Gibbons and K. Maeda, Nucl. Phys.B, {\bf 298}, 741 (1988).
\bibitem{spin2}Y.M. Cho, Phys.Rev.Lett., {\bf 68}, 3133 (1992).
\bibitem{quant}E.W. Kolb, D. Salopek and M.S. Turner, Phys.Rev.D, {\bf 42}, 3925 (1990).
\bibitem{third}W. Buchmuller and N. Dragon, Nucl. Phys., {\bf B321}, 207 (1989);
J.A. Casas, J. Garcia-Bellido and M. Quiros, Class. Quant. Grav., {\bf 9}, 1371 (1992).
\bibitem{brans}C.H. Brans, Class. Quant. Grav., {\bf 5}, L197 (1988).
\bibitem{magnano}G. Magnano and L.M. Sokolowski, Phys.Rev.D, {\bf50},5039 (1994).
\bibitem{kinf1}M. Sami Mod. Phys. lett., {\bf A18}, 691 (2003);
A. Feinstein, Phys.Rev.D, {\bf 66}, 063511 (2002);
\bibitem{underwood}B. Underwood, Phys.Rev.D, {\bf 78}, 023509 (2008).
M. Fairbin and M. Tytgat, Phys.Lett.B, {\bf 46}, 1 (2002);
\bibitem{kinfprob}L. Kofman and A. Linde, JHEP, {\bf 0207}, 004 (2002).
\bibitem{recent}R. C. de Souza and G. M. Kremer, arXiv:0809.2331 [gr-qc].
 
\end{thebibliography}
\end{document}